\documentclass[%
 reprint,
superscriptaddress,
 amsmath,amssymb,
 aps,
]{revtex4-2}

\usepackage{graphicx}
\usepackage{dcolumn}
\usepackage{bm}
\usepackage{hyperref}
\usepackage{xcolor}

\begin{document}

\preprint{APS/123-QED}

\title{Informational Memory Shapes Collective Behavior in Intelligent Swarms}

\author{Shengkai Li}
\thanks{These two authors contributed equally.}
\affiliation{Department of Physics, Princeton University, Princeton, NJ 08544, USA}%

\author{Trung V. Phan}%
\thanks{These two authors contributed equally.}
\affiliation{%
Department of Natural Sciences, Scripps and Pitzer Colleges, \\ Claremont Colleges Consortium, Claremont, CA 92110, USA}%

\author{Luca Di Carlo}
\affiliation{Department of Physics, Princeton University, Princeton, NJ 08544, USA}%

\author{Gao Wang}
\affiliation{
Wenzhou Institute, University of Chinese Academy of Sciences, Wenzhou, Zhejiang 325000, China}%

\author{Van H. Do}
\affiliation{%
Homer L. Dodge Department of Physics and Astronomy, University of Oklahoma, 440 W. Brooks St. Norman, OK 73019, USA}%

\author{Elia Mikhail}
\affiliation{%
Department of Electrical and Computer Engineering, Princeton University, Princeton, New Jersey 08544, USA}%

\author{Robert H. Austin}
\thanks{austin@princeton.edu}
\affiliation{%
Department of Physics, Princeton University, Princeton, NJ 08544, USA}%

\author{Liyu Liu}
\thanks{liuliyu@fudan.edu.cn}
\affiliation{%
Human Phenome Institute, Fudan University, Shanghai 201203, China}%

\date{\today}

\begin{abstract}
{We present an experimental and theoretical study of 2-D swarms in which collective behavior emerges from both direct local mechanical coupling between agents and from the exchange and processing of information between agents. Each agent, an air-table drone endowed with internal memory and a binary decision variable, updates its state by integrating a time series of memories of local past collisions. This internal computation transforms the drone swarm into a dynamical information network in which history-dependent feedback drives spontaneous complete spin polarization, pitchfork bifurcated spin collectives, and chaotic switching between collective states. By tuning the depth of memory and the decision algorithm, we uncover a memory-induced phase transition that breaks spin symmetry at the population level. A minimal theoretical model maps these dynamics onto an effective potential landscape sculpted by informational feedback, revealing how temporally correlated computation can replace instantaneous forces as the driver of collective organization, informed by experiments. These results position physically interacting drone swarms as a model system for exploring the physics of informational drone ensembles whose emergent behavior arises from the interplay between physical interaction and information processing.}
\end{abstract}


\maketitle

\begin{figure*}[ht!]
\centering
\includegraphics[width=0.95\textwidth,keepaspectratio]{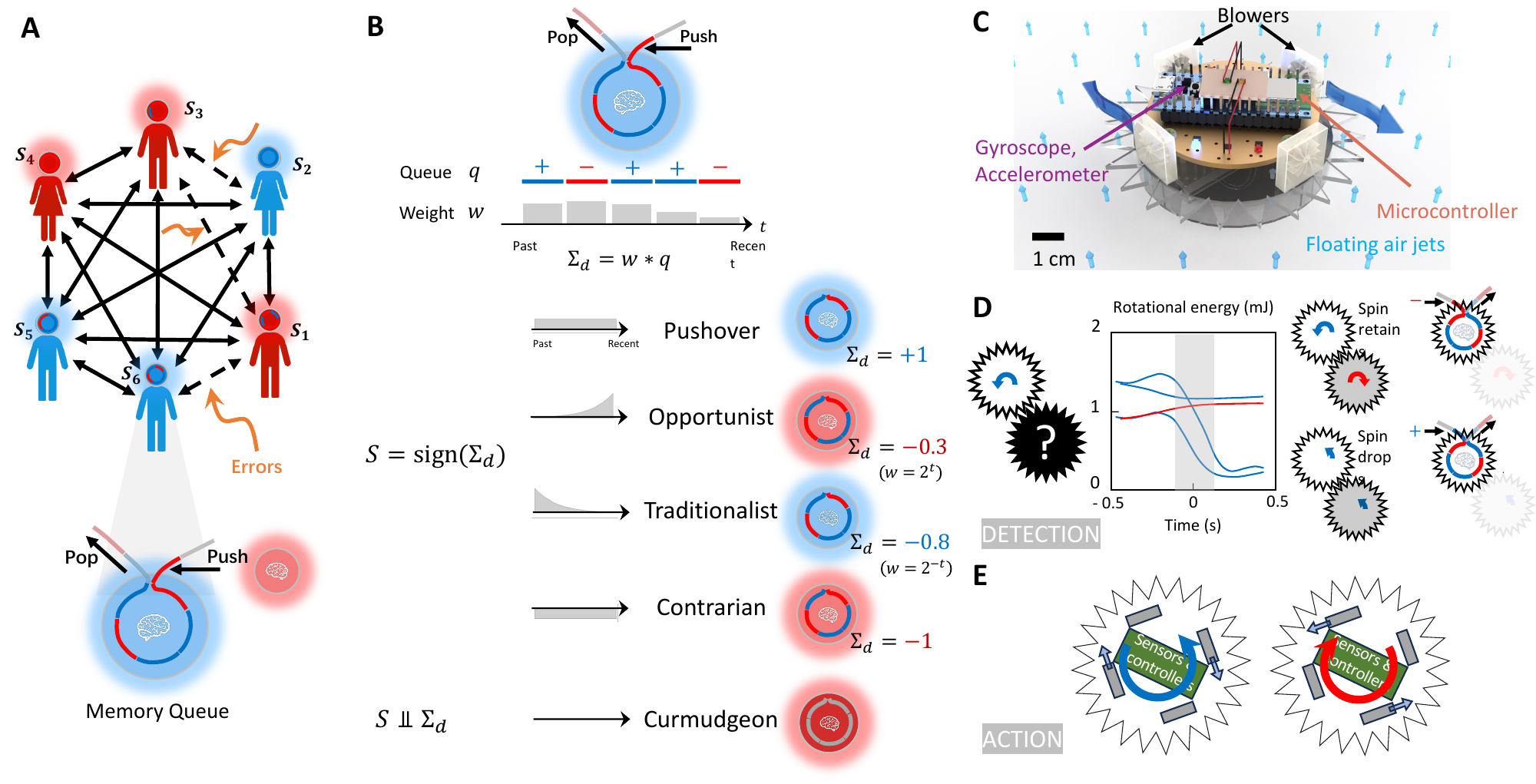}
\caption{\textbf{Dynamics of opinion exchange.} \textbf{A.} People change their opinions after evaluating opinions from the others. The evaluation depends on a finite memory of recent observations of the peer stored in a queue. \textbf{B.} The state updates with the weighted memory $\Sigma_d=\sum_i w_i q_i$. Observations $q_i\in\{+1,-1\}$ are weighted by $w_i$ depending on their personalities. One common decision is to follow the majority where $w_i$ is constant such that $S=\text{sign}(\Sigma_d)$ picks the larger counts of the spin. Similarly, negative constant weight defines an agent moving against the majority, and higher weights on the more recent events defines an opportunist. Curmudgeons with strong systematic biases have independence of weighted memory. \textbf{C.} A physical agent (spinner) has a microcomputer with gyroscope, accelerometer, and actuates the state of the four blowers. \textbf{D.} A spinner records information of surrounding spinners through mechanical interactions (gray shades) that same spins drop spin upon collision while opposite spins do not \cite{spinners}. \textbf{E.} A spinner can choose to spin counterclockwise or clockwise by selecting appropriate blowers depending on its internal algorithm. See \href{https://www.dropbox.com/scl/fi/ech59xy6nexs2m2xzrlow/SI1_smartSpinners_c.mp4?rlkey=4ks7wko1lo4v6oxdx0ewqob45&dl=0}{SI1.mp4} for demonstration.}
\label{concept}
\end{figure*}

\textit{Introduction} - Many living systems exhibit behaviors shaped by past physical interactions, with memory depth playing a crucial role. Birds align with neighbors to form flocks~\cite{vicsek1995novel}, bacteria integrate chemical signals to navigate~\cite{segall1986temporal,gosztolai2020cellular}, and humans vote or invest based on memories~\cite{lacy2017votes}. Such responses extend beyond immediate stimuli and involve internal information processing. Memory-based feedback can generate complex behaviors, such as nonreciprocal interactions~\cite{reciprocal}. Memory-influenced opinion dynamics exemplifies this and has attracted a sustained interest in physics and mathematics~\cite{galam2008sociophysics,axelrod,castellano2009statistical,lorenz,evolution,breslauer} (Fig.~\ref{concept}A), with political polarization shaped by how people recall and evaluate past events~\cite{axelrod2021preventing}.

Here we work with drone agents that autonomously control their rotation and update a binary spin state interpreted as opinion~\cite{yu2023programmable,vansaders2023informational,chen2024emergent,li2021programming}. Agents perceive the local spin of colliding partners and modify their own spin based on internal memory and decision rules, reflecting basic ``personalities'' such as conformity, contrarianism, or stubbornness~\cite{galam2023unanimity} (Fig.~\ref{concept}A). Although physical collisions follow deterministic laws~\cite{spinners}, agent responses are driven by informational feedback accumulated over time. Personality remains fixed (Fig.~\ref{concept}B), but memory processing enables spontaneous polarization, oscillations, and chaotic transitions at the collective level~\cite{lynn2019surges,silverberg2013collective,te2020effects}.

{This system exemplifies how sensing, actuation, and memory together generate adaptive collective behavior \cite{kaspar2021rise}.
Unlike many active matter studies with fixed chirality, our agents use information-driven, memory-based interactions linking past encounters to macroscopic order, offering a physically interpretable model of informational feedback.
The results advance non-reciprocal dynamics \cite{reciprocal,chen2024emergent,brandenbourger2019non}, complement social-polarization models \cite{galam2008sociophysics,axelrod}, and extend chiral-ensemble studies \cite{scholz2021surfactants,scholz2018rotating,yang2020robust,tan2022odd} to systems whose spins are set by internal states. Earlier works examined memory in collective dynamics: opinion-formation models integrate all past neighbors without cutoff \cite{zhang2018opinion}; robotic active matter uses passive sensorial delay \cite{mijalkov2016engineering}; and hydrodynamic feedback yields emergent chirality reversal \cite{chen2017weak}.
Here, each agent retains a tunable, finite record of recent interactions, allowing controlled responsiveness and stability.
This finite, programmable memory serves as an internal degree of freedom that governs symmetry breaking and collective organization.}\\

\textit{Embedding Computation, Memory, Personality and Communication - } The drones are driven spinning gears floating on an air table, with an onboard Arduino microcomputer, sensors, and blowers to direct tangential air flow. Among the sensors, the accelerometer detects collision events while the gyroscope determines spin changes. Depending on the information the microcomputer gets from the sensors and the history of previous collisions and how it interprets the past events, each spinner makes a decision to possibly alter its intrinsic spin by actuating the blowers to change their spin (see Fig.\ref{concept}).

The ability to recognize another agent’s spin direction arises from the physics of collisions between spinning and translating bodies \cite{spinners}, akin to antiferromagnetism \cite{guo2023non}: Collisions between same-handed spinners reduce both spins, while opposite-handed ones retain their spins. Each spinner has four blowers arranged so that activating one pair produces counterclockwise ($+$) motion, and the other pair produces clockwise ($-$) motion.

When a collision is detected via the accelerometer and the colliding agent’s spin is inferred from the gyroscope, the result is pushed onto an $M$-bit queue. Due to complex collision dynamics, spin determination is imperfect; the current setup achieves $70~\%$ accuracy. Thus, the stored sequence of $+1$'s and $-1$'s, representing past collisions, is not exact.

The memory depth $M$ is set before each experiment, ranging from $1$ (only the previous collision stored) to $21$. The queue operates first-in/first-out. The spinner sets its spin based on the weighted sum of the queue $\Sigma_d = \sum_{i=1}^M w_i q_i$, where $q_i$ is the $i$th spin and $w_i$ its weight. This convolution of memory and kernel resembles how bacteria assess chemoattractant gradients over time \cite{segall1986temporal,gosztolai2020cellular}. Like bacteria choose to run or tumble, our spinner chooses spin direction via $s=\text{sign}(\Sigma_d)$, a nonlinear feedback that may enhance polarization \cite{leonard2021nonlinear}.

For example, a pushover spinner sets all weights $w_i = 1$ and aligns its spin with the sum of the past $M$ events, seeking consensus. A curmudgeon remains fixed at $+$ or $-$ regardless of events. An opportunist favors recent events, while a traditionalist emphasizes the most distant ones. A contrarian sets its spin opposite to the majority in its memory.

\begin{figure*}[!thb]
\centering
\includegraphics[width=0.94\textwidth]{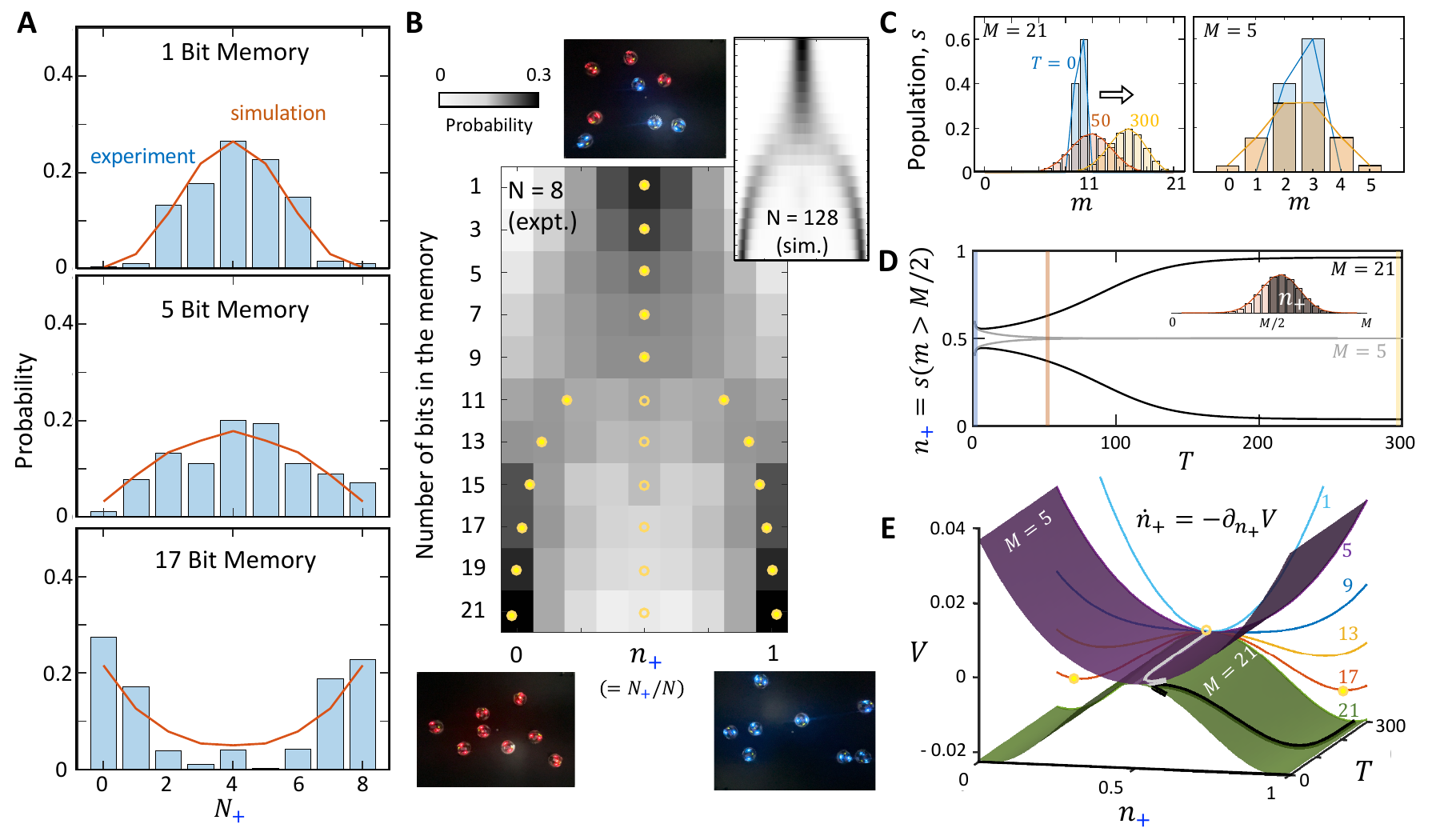} 
\caption{\textbf{Memory-induced spontaneous symmetry breaking.}  \textbf{A.} Probability of states for a collective of pushover with each building its memory of the past $M$ collision inference and following the majority in memory for different $M$. \textbf{B.} When the memory size is small, there is no net polarization. When the size of memory is sufficiently large, the population collapses where one spin dominates the other. See \href{https://www.dropbox.com/scl/fi/00hoxet5xt789koj18ohv/SI2_pushover_c.mp4?rlkey=tfi2tcqqw74h9zx5tbwf2il16&dl=0}{SI2.mp4} for experiment videos. The inset shows the simulation result for large N (=128). See \href{https://www.dropbox.com/scl/fi/468uppenhiqm2em1i1ajj/SI7_simulationBigN_c.mp4?rlkey=qxzq3nsvyvrx7e1lcxjuv95y3&dl=0}{SI7.mp4} for the simulation videos. \textbf{C.} Population over time for spinners with $m$ bits of $+$. When the initial population of $+$ spinners is slightly higher than the $-$ spinners, memory size $M=21$ attracts the population to higher $+$ memory states while $M=5$ erases the initial bias and leads the collective to an even distribution of $+$ and $-$. \textbf{D.} The fraction of $+$ over time for $M=5$ and $M=21$. Here the detection error $\eta=0.3$. \textbf{E.} The fraction of $+$ spinners $n_+=N_+/N$ follows the gradient of an effective potential $V$ which varies from a single well when $M$ is small to a double well when $M$ exceeds a critical value.}
\label{theory}
\end{figure*}

Although each spinner’s state is binary ($+$ or $-$), its detailed memory affects its response to collisions. For instance, a $5$-bit $+$ pushover with $(3+,2-)$ is more likely to change than one with $(5+,0-)$ (a deep $+$ state). We later show how the evolution of these finer substates influences emergent consensus (Fig.~\ref{eq:potential}). Our agents decide future spin based on past experiences, enabling highly non-Markovian behavior, diverging from purely physics-based systems. For example, pushovers align with the perceived majority spin, even if sampled with error, resembling human information processing \cite{lynn2020human,jiang2023neurocomputational}.\\

\textit{Memory induced spontaneous symmetry breaking of the pushovers - } Starting with equal numbers of $+$ and $-$ spinners, spontaneous symmetry breaking can occur due to nonlinear interactions \cite{guralnik}. While initial spin polarization is zero, its evolution depends on decision rules, memory depth, and noise. Notably, the progression of spontaneous polarization is strongly influenced by the memory depth $M$. In experiments with $M=1$, the collective stays around 50/50 (Fig. \ref{theory}).
But with larger memory, for instance $M = 17$, initially unpolarized pushover spinners settles into either all $+$ or all $-$ states, each with 50\% probability depending on the initial conditions, indicating spontaneous symmetry breaking. A sharp transition from no-polarization to complete symmetry breaking occurs around $M \sim 9$ (Fig.\ref{theory}), resembling tipping points in complex systems, such as ecological transitions with early warning signals \cite{xu2023non}.

To build an intuition for spontaneous symmetry breaking, we studied how spinner populations with different memory configurations evolve over time. The current population fraction of spinners with $m$ bits of $+$ in memory, $s_m$ where $0\le m \le M$, is contributed by three populations in the past: (1) Spinners previously with $m$ bits of $+$ and did not change their state, (2) Spinners previously with $m-1$ bits of $+$ and then detected another $+$ spinner, (3) Spinners previously with $m+1$ bits of $+$ and detected another $-$ spinner.

As an example, we apply the above to a collective started with an initial population of $+$ spinners slightly higher than the $-$ spinners. With memory size $M=21$, the population is attracted to higher $+$ memory states. In contrast, when $M=5$, the evolution of the population erases the initial bias and leads the collective to an even distribution of $+$ and $-$ (Fig.\ref{theory}C,D). {By tracking the transition of memory configurations (strings of $+/-$) of the spinners (Sec.I.A in \href{https://www.dropbox.com/scl/fi/idzlhmardmp4v08nmil4w/revised_SI.pdf?rlkey=aqxm4ga6ps5s0am5cll75jr0x&dl=0}{SI}), we can show that} the fraction of $+$ spinners, $n_+=N_+/N$ where $N_+$ and $N$ are the numbers of $+$ spinners and total spinners, follows the gradient of an effective potential well $V$ with $\dot{n}_+=-{\frac{1}{\tau}}\frac{\partial V}{\partial n_+}$ where {$\tau$ is the average state updating time} and the potential well is shaped by the memory size $M$ and the detection error $\eta$ as
\begin{eqnarray}
V=\frac{1}{2}\left[1-(1-2\eta)\sqrt{\frac{2M}{\pi}}\right]\left(n_+-\frac{1}{2}\right)^2+O(n_+^4)\label{eq:potential}
\end{eqnarray}
{where the $O(n_+^4)$ expansion is $\frac{(1-2\eta)^3}{3\sqrt{2\pi}} M^{3/2}(n_+-0.5)^4$.}

{As the memory size $M$ increases, the sign of the quadratic term changes from positive to negative, transforming the potential from a single well to a double well and leading to a supercritical pitchfork bifurcation \cite{strogatz}}. Lower detection error $\eta$ also promotes this transition. The critical memory $M_c = \pi/2(1 - 2\eta)^2$ yields $9.8$ bits for the 30\% error rate observed experimentally, matching the transition point. A scan of symmetry breaking, via variance $\sum_i (n_{+,i} - 0.5)^2 p_i$, confirms this criticality (Sec. I.A in \href{https://www.dropbox.com/scl/fi/idzlhmardmp4v08nmil4w/revised_SI.pdf?rlkey=aqxm4ga6ps5s0am5cll75jr0x&dl=0}{SI}). The memory–error dependence resembles polarization in opinion-dynamics models \cite{axelrod2021preventing}; unlike Axelrod’s agents, which form simultaneous opposing clusters, our spinners occupy one extreme at a time and collectively switch. Simulations with larger $N=128$ confirm a bifurcation near $M=10$ (Fig.~\ref{theory}).

The above demonstrates non-Markovian effects in systems with evolving internal states \cite{wang2022robots,barkan2023multiple}.
In the Markovian limit $M=1$, reciprocal interactions yield equipartition of microstates and a binomial distribution for $N_+$ (Sec. I.B in the \href{https://www.dropbox.com/scl/fi/idzlhmardmp4v08nmil4w/revised_SI.pdf?rlkey=aqxm4ga6ps5s0am5cll75jr0x&dl=0}{SI}
). Increasing $M$ strengthens non-Markovianity, producing qualitatively new collective behavior beyond a simple timescale effect.
Symmetry breaking persists under error, and tolerance rises with memory depth.

Another perspective on polarization views it as a limit of reliable information transmission \cite{shannon1948mathematical,chong2023multilegged}.
Each collision transmits one noisy bit about the collective state through a binary symmetric channel with error $\eta$ and Shannon capacity $C=1-H(\eta)$, where the entropy $H(\eta)=-\eta\log_2\eta-(1-\eta)\log_2(1-\eta)$.
An agent with memory depth $M$ acts as a length-$M$ repetition encoder with majority decoding, giving rate $R=1/M$.
Shannon’s theorem requires $R<C$, yielding $M_c\simeq1/[1-H(\eta)]$.
For $\eta=0.3$, this predicts $M_c\approx8.4$, close to the observed transition near $9$.
Expanding $H(\eta)$ near $\eta=0.5$ gives the same scaling $M_c\propto(0.5-\eta)^{-2}$ as the state-transition theory, showing that symmetry breaking occurs when information transmission saturates channel capacity.\\
\color{black}

\begin{figure}[!thb]
\centering
\includegraphics[width=0.48\textwidth]{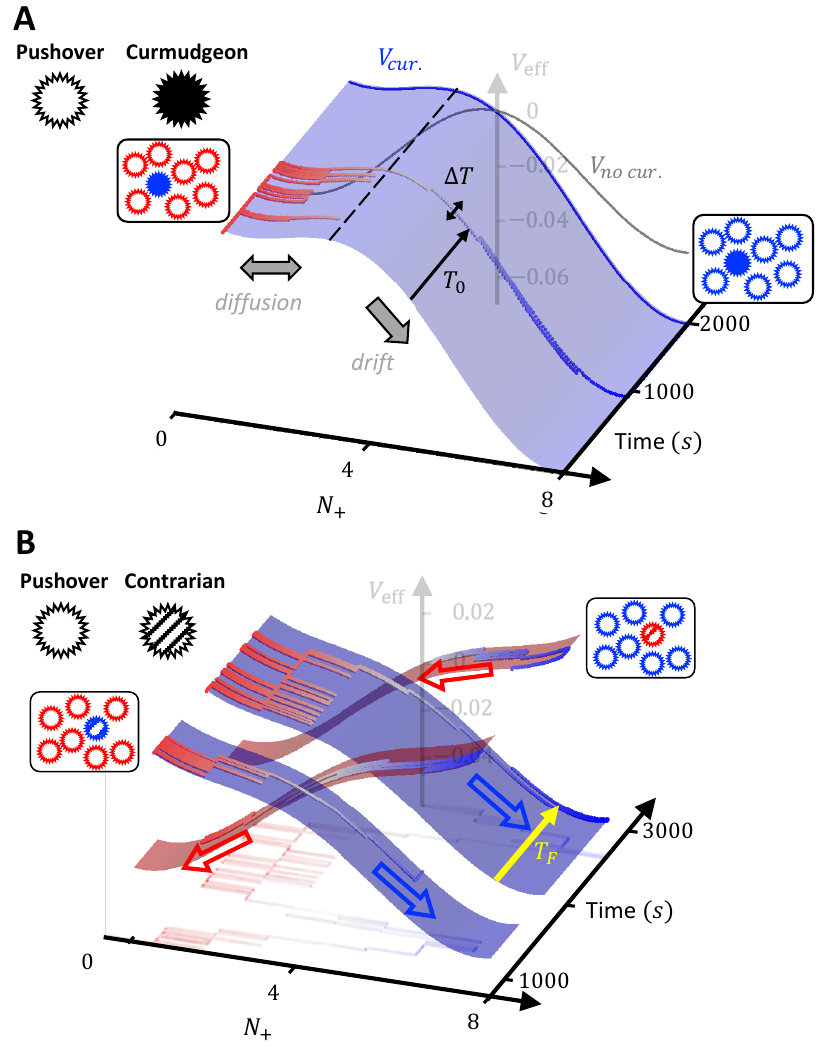} 
\caption{\textbf{Curmudgeon and contrarian.} \textbf{A.} Among pushover spinners with 27-bit memories (above critical), a curmudgeon has a constant opinion which is able to distort the symmetric double-well potential ($V_{\text{no cur.}}$) of the pushovers to a biased double-well ($V_{\text{cur.}}$). An experiment where a group of pushovers with all $-$ bits get directed by a $+$ curmudgeon after the initial diffusion stage. See \href{https://www.dropbox.com/scl/fi/hwpurnhzm82t02a6za2si/SI3_curmudgeon_c.mp4?rlkey=swweg6795i8kwuziapp35l5q6&dl=0}{SI3.mp4} for experiment video.  \textbf{B.} A contrarian always acting the opposite to the majority acts like a curmudgeon on a short time scale. Here $M=17$. See \href{https://www.dropbox.com/scl/fi/b14g365dorr8g0krqy7i2/SI4_contrarian_c.mp4?rlkey=sxb7bjo0zoituq4odzi8t0f75&dl=0}{SI4.mp4} for experiment video.}
\label{dd}
\end{figure}

\textit{A curmudgeon among the pushovers - }A curmudgeon agent follows its own fixed opinion \cite{galam2024fake} and responds nonreciprocally \cite{reciprocal} to collisions, unlike pushovers that react based on memory. Using Bluetooth, we changed curmudgeons’ chirality to demonstrate their influence—pushovers adjusted accordingly (\href{https://www.dropbox.com/scl/fi/hwpurnhzm82t02a6za2si/SI3_curmudgeon_c.mp4?rlkey=swweg6795i8kwuziapp35l5q6&dl=0}{SI3.mp4} and Fig. S7 in \href{https://www.dropbox.com/scl/fi/idzlhmardmp4v08nmil4w/revised_SI.pdf?rlkey=aqxm4ga6ps5s0am5cll75jr0x&dl=0}{SI}). Adding curmudgeons tilts the double-well potential (Eq.\ref{eq:potential}); even 12.5\% (1 of 8 spinners) significantly biases the system when $M=27$ (Fig.\ref{dd}A, Sec. I.C in \href{https://www.dropbox.com/scl/fi/x6qq80cppa3dq7yc169ew/Spinner_SI.pdf?rlkey=pc8c3gx4ciwl0vh61jqhk7m23&dl=0}{SI}). {Starting from all spinners in the $-$ state, a single $+$ curmudgeon seeds a field-biased, nucleation-like switch to $+$. Unlike spontaneous classical nucleation \cite{cates2023classical}, here we have the curmudgeon as a nucleation source. With both $+$ and $-$ curmudgeons present, the larger subpopulation typically wins over time; for equal numbers, the system undergoes a pitchfork bifurcation at a higher critical memory (Sec.\,IV.D, \href{https://www.dropbox.com/scl/fi/idzlhmardmp4v08nmil4w/revised_SI.pdf?rlkey=aqxm4ga6ps5s0am5cll75jr0x&dl=0}{SI}). If curmudgeons randomly flip sign, they act as dissenters in modified Vicsek model and raise the consensus threshold (Sec.\,IV.E, \href{https://www.dropbox.com/scl/fi/idzlhmardmp4v08nmil4w/revised_SI.pdf?rlkey=aqxm4ga6ps5s0am5cll75jr0x&dl=0}{SI}).}

Interestingly, when memory is below the critical threshold ($M < M_c$), pushovers are not affected by curmudgeons and remain near a 50/50 split. To influence others, a curmudgeon needs a crowd that highly values peer opinions. At very large $M$, the crowd's shift toward the curmudgeon’s state happens abruptly \cite{lynn2019surges}, due to slow diffusion across the barrier followed by rapid descent along a steep potential (Fig.~\ref{dd}A). This timescale gap widens with $M$, as diffusion time $T_0$ grows with memory size: $M^2 = \langle m(T_0)^2 \rangle \propto T_0$.\\ 

\textit{A contrarian among the pushovers - } A contrarian sets its spin opposite the majority in its memory \cite{galam2004contrarian}, often standing alone like a curmudgeon. As the crowd shifts toward it, new interactions reveal the change and it flips again, sustaining opposition. Experiments (Fig.~\ref{dd}B) show a small phase delay from detection–response time to demographic changes.

The flipping interval $T_F$ varies, with short intervals more common than long ones. Long $T_F$ follows an exponential distribution, while short $T_F$ approximates a power law (Fig. S8 in \href{https://www.dropbox.com/scl/fi/idzlhmardmp4v08nmil4w/revised_SI.pdf?rlkey=aqxm4ga6ps5s0am5cll75jr0x&dl=0}{SI}), reflecting the time needed for a contrarian to overwrite its memory (about $M$ collisions). This drives the alternation between consensus states. The effect is more pronounced with high memory (Fig. S10 in \href{https://www.dropbox.com/scl/fi/idzlhmardmp4v08nmil4w/revised_SI.pdf?rlkey=aqxm4ga6ps5s0am5cll75jr0x&dl=0}{SI}), which strengthens attraction to each consensus well and accelerates transitions, as seen in curmudgeon experiments.

A minimal model coupling the contrarian and pushovers captures key dynamics. Pushovers influence the contrarian’s memory: $\dot{c} = k_c p + \xi_c$, where $p$ is the excess $+$ population and $c$ the excess $+$ bits in the contrarian. The contrarian, in turn, drives the population: $\dot{p} = -k_p \text{sgn}(c) + \xi_p$. Coupling these gives
\begin{eqnarray}
\ddot{c}=-k~\text{sgn} (c) + \xi \label{eq:contrarian}
\end{eqnarray}
where $\xi=\xi_p/k_p+\xi_c/k_p k_c$ and $k=k_c k_p$. This interaction produces back-and-forth consensus switching analogous to stick-slip dynamics. Large memory $M$ causes sharp transitions, whereas intermediate ‘confused’ states yield rapid flips. The flipping interval follows a power law (exponent $\approx -1.5$) for short times, with an exponential tail scaling as $M^2$ \cite{yan2024avalanches}. Large $M$ allows rare, long intervals that sustain demographic switching (Sec. I.D in \href{https://www.dropbox.com/scl/fi/7eitx8y4hgg7k7dwmxcbi/revised_SI.pdf?rlkey=7dpgyerr5grr2exg5t10illnd&dl=0}{SI}).

{The population oscillation that arises with a contrarian resembles the swap phase reported in nonreciprocal systems\cite{fruchart2021non}. In our model, the memory depth $M$ controls this behavior: for small $M$, the population remains homogeneous and the contrarian has little influence on the pushovers, whereas increasing $M$ generates the swap phase. This underscores that memory depth, the key ingredient of this work, governs the collective dynamics of memory-bearing agents.\\}

\begin{figure}[ht!]
\includegraphics[width=0.48\textwidth,keepaspectratio]{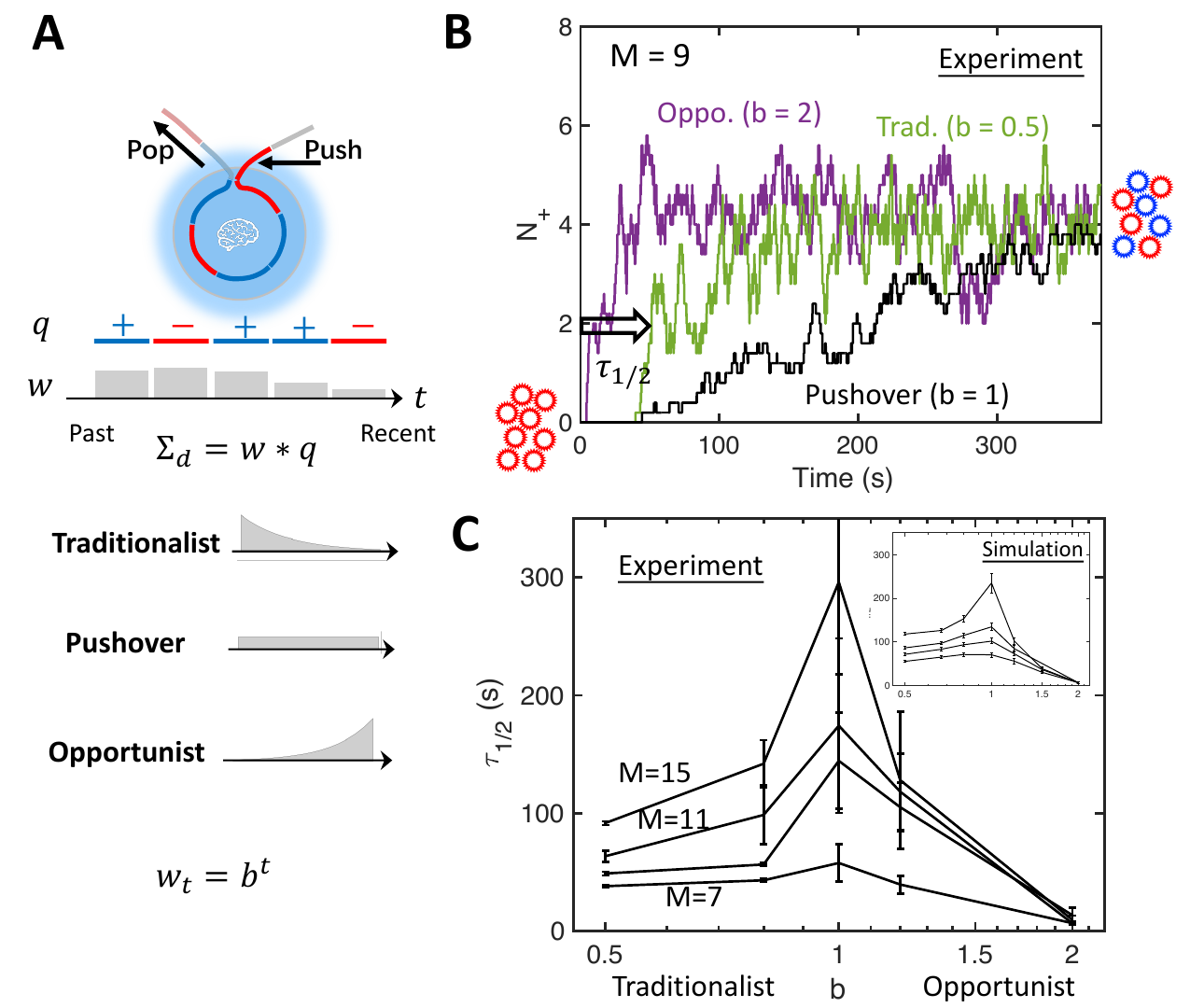}
\centering
\caption{\textbf{Weighted memory.} \textbf{A.} The weight on the memory $w\propto b^t$ shifts its focus from the past to the recent as the weight bias $b$ increases from $0$. \textbf{B.} The number of $+$ spinner, $N_+$, increases with time after all spinners start with pure $-$ bits. The speed of the process varies among traditionalist, pushover, and opportunist. See \href{https://www.dropbox.com/scl/fi/rwmdfpkitif6zy6stkvct/SI5_timeWeight_c.mp4?rlkey=v3mprpgx6spa0uzwjluknybgo&dl=0}{SI5.mp4} for experiment video. \textbf{C.} Half-lives to reach 50/50 for different bias ($b$) and memory size ($M$) combinations from simulation.}
\label{weightedMemory}
\end{figure}

\textit{The traditionalist, pushover, and opportunist.} - Humans weigh past and recent memories differently when making decisions \cite{ebbinghaus1885gedachtnis,rubin1999precise,wozniak1995two}. We explore how exponential memory weight, $w \propto b^t$, influences opinion polarization. For $b=1$, spinners are pushovers; $b>1$ emphasizes recent events (opportunists) while $b<1$ favors distant ones (traditionalists). The same memory can lead to opposite decisions depending on the weighting (Fig.~\ref{concept}B). Starting all spinners with $-$ spin and $-$ bits, we measure the time to reach an even distribution (Fig.~\ref{weightedMemory}B,C). We find: 1. Traditionalists respond more slowly than opportunists; 2. Response time grows with memory $M$, with a sharp rise beyond $M_c$, consistent with earlier results; 3. The peak occurs at $b=1$ (pushover); {4. The above hold for larger $N$ (Sec.IV.F in \href{https://www.dropbox.com/scl/fi/idzlhmardmp4v08nmil4w/revised_SI.pdf?rlkey=aqxm4ga6ps5s0am5cll75jr0x&dl=0}{SI})}.

Intuitively, the opportunist’s discount for older memory shortens effective memory and weakens symmetry breaking. Although weighting distant events seems to enhance the breaking, experiments and simulations show the opposite because outdated information dominates. In the limit $b \ll 1$, traditionalists rely on a single bit from $M$ collisions earlier, behaving as originalists \cite{solum2018originalism}.\\ 

\textit{Conclusion - } {Our system is an ensemble where  sensing, memory, and complex response yield adaptive collective behavior \cite{kaspar2021rise}. Unlike active matter driven by instantaneous forces or fixed design, agents here self-organize through information-driven dynamics that couple interaction history to motion, producing effective nonreciprocal interactions \cite{reciprocal,chen2024emergent,brandenbourger2019non}. The phenomena emerged experimentally first. Theory and simulation were then developed to rationalize and quantify them.} 

{Using tools from statistical physics, we map collective opinion dynamics to a potential landscape, yielding a quantitative framework for memory-driven symmetry breaking. This extends chiral collective dynamics \cite{scholz2021surfactants,scholz2018rotating,yang2020robust,tan2022odd} to systems where handedness and coordination are emergent computational states, and complements social-polarization models \cite{galam2008sociophysics,axelrod}. Although demonstrated with air-table drones, the principles are platform-independent and applicable to biological collectives. Our results identify memory as a physical degree of freedom in nonequilibrium systems where past interactions shape future behavior and enable studies of memory-driven collectives that sense, learn, and react \cite{miskin2020electronically,lassiter2025microscopic}.}\\

\textit{Acknowledgement - } {We thank Stanley H. Chidzik for PCB manufacture,
Huaicheng Chen for the tracking code, Michael te Vrugt
for valuable discussions at the early stage of this research,
Robert Axelrod, Jie Bao, Ruihua Fan, Daniel Goldman, Stanley Nicholson, Milo\v{s} Nikoli\v{c}, Huy D. Tran, and
Bryan VanSaders for helpful discussions. We thank Jakob M\"{o}llmann for correcting typos and errors. The authors acknowledge the very helpful and valuable advice from the referees.} This work was supported by the Center for Physics of Biological Function at Princeton University and the National
Natural Science Foundation of China (Nos.T2350007, 12404239, 12174041).


\bibliography{apssamp}

\end{document}